\documentclass[12pt]{article}
\usepackage[english]{babel}
\usepackage{graphicx}
\usepackage{bm}
\usepackage{setspace}
\usepackage{varioref}
\usepackage{booktabs}
\usepackage{rotating}
\usepackage{xfrac}
\usepackage{adjustbox}
\usepackage{longtable} 
\usepackage{ragged2e}
\usepackage[osf,sc]{mathpazo} 
\usepackage{microtype} 
\usepackage[left=0.93in,right=0.93in,bottom=0.93in,top=0.93in]{geometry} 
\usepackage[hang, small,labelfont=bf,up,textfont=it,up]{caption} 
\usepackage{booktabs} 
\usepackage{float} 
\usepackage[colorlinks=false]{hyperref}
\hypersetup{colorlinks, citecolor=blue, urlcolor=blue, linkcolor=blue} 
\usepackage{lettrine} 
\usepackage{paralist} 
\usepackage{enumitem} 
\usepackage{amsfonts}
\usepackage{amsmath}
\usepackage{amsthm}
\usepackage{mathtools}
\usepackage{hyperref}
 
\theoremstyle{definition}

\usepackage{subcaption} 
\usepackage{abstract} 
\usepackage{url}
\usepackage{titlesec} 
\titleformat{\section}[block]{\large\scshape\centering}{\thesection.}{1em}{} 
\titleformat{\subsection}[block]{\large}{\thesubsection.}{1em}{} 
\usepackage{fancyhdr} 

\usepackage{booktabs}
\usepackage{pdflscape}

\usepackage[english]{isodate}

\usepackage[longnamesfirst]{natbib}

\def\sym#1{\ifmmode^{#1}\else\(^{#1}\)\fi}
\usepackage{rotating,booktabs}

\usepackage{xcolor} 
\usepackage{placeins}
\usepackage{cleveref}
\begin{document}

\makeatletter
\renewcommand*{\@fnsymbol}[1]{\ensuremath{\ifcase#1\or \star \or \dagger\or \ddagger\or
    \mathsection\or \mathparagraph\or \|\or **\or \dagger\dagger
    \or \ddagger\ddagger \else\@ctrerr\fi}}
\makeatother

\title{Inference for Local Projections%
\hspace{2pt}%
\thanks{%
We thank the Royal Economic Society for the invitation to participate in the 2024 meetings in Belfast. Comments and suggestions from Raffaella Giacomini, Jaap Abbring, and conference participants helped us improve the paper. The views expressed in this paper are the sole responsibility of the authors and do not necessarily reflect the views of the Federal Reserve Bank of San Francisco or the Federal Reserve System.%
}%
}

\author{%
Atsushi Inoue\hspace{0.5pt}%
\thanks{Vandebilt  University (\href{mailto:atsushi.inoue@vanderbilt.edu}{atsushi.inoue@vanderbilt.edu}).}%
\and \`{O}scar Jord\`{a}\hspace{0.5pt}%
\thanks{Federal Reserve Bank of San Francisco; and Department of Economics, 
University of California, Davis (\href{mailto:oscar.jorda@sf.frb.org}{oscar.jorda@sf.frb.org}; \href{mailto:ojorda@ucdavis.edu}{ojorda@ucdavis.edu}) and CEPR.}%
\and Guido M. Kuersteiner\hspace{2pt}%
\thanks{Department of Economics, 
University of Maryland (\href{mailto:gkuerste@umd.edu}{gkuerste@umd.edu}).}%
}

\maketitle
\thispagestyle{empty}

\small


\begin{abstract}
\small
\singlespacing
\noindent

Inference for impulse responses estimated with local projections presents interesting challenges and opportunities. Analysts typically want to assess the precision of individual estimates, explore the dynamic evolution of the response over particular regions, and generally determine whether the impulse generates a response that is any different from the null of no effect. Each of these goals requires a different approach to inference. In this article, we provide an overview of results that have appeared in the literature in the past 20 years along with some new procedures that we introduce here.


\bigskip

\end{abstract}

\allowbreak \newpage

\newpage

\clearpage
\quad

\pagenumbering{arabic}

\section{Introduction}

Impulse responses are often used to characterize dynamic systems. When the analyst wants to remain agnostic about the data generating process (DGP), impulse responses are often estimated using the method of local projections proposed by \cite{Jorda2005} or LPs for short. LPs consist of projecting future outcome variables on current and past information on the intervention (or impulse), the outcome and other exogenous variables. In other words, they are simple regressions with a particular dynamic error structure. 

This article discusses how to conduct inference for LPs with an emphasis on general principles. Relative to vector autoregressions (VARs), LPs generally have lower bias but higher variance though in infinite samples, when the lag structure grows with the sample, both methods generate the same impulse response and similar parameter estimation uncertainty \citep[see, e.g.][]{PMWolf2021, Xu2023}. We organize our discussion around three main topics: (1) point-wise inference; (2) simultaneous inference; and (3) significance. 

Point-wise inference, by far the most popular, refers to the uncertainty with which individual parameters of the impulse response are estimated. It is presented graphically by means of error bands. Simultaneous inference refers to the uncertainty about sets of impulse response coefficients. It can be presented graphically as error bounds that accommodate families of hypotheses. Because these bounds are generally wider than error bands, practitioners do not generally show them. Both point-wise and simultaneous inference are based on the Wald principle.

In this paper, we introduce a new concept: \emph{significance bands}. These bands refer to the null hypothesis that the impulse does not generate a response, or in the parlance of experiments, that the policy intervention has no effect. The Lagrange-Multiplier (LM) principle considerably simplifies the construction of significance bands, which are a useful complement to the practice of presenting error bands since many researchers are often concerned about the absence of a response to an impulse.

We set the stage with a brief introduction to LPs to highlight where the properties of the estimated response come from and how they affect inference. Next, we discuss point-wise inference. We highlight several recent developments in the literature, including feasible generalized least-squares methods (LP-FGLS), lag-augmentation (LP-LA), and with a brief mention of Bayesian inference (BLP).

Thinking about simultaneous inference requires system estimation of LPs. Thus we begin by framing such system estimation within the class of generalized method of moments (GMM) estimators. This allows us to cover instrumental variable estimation. With this scaffolding in place, we review two important papers in this arena, \cite{Jorda2009} and \cite{MontielOleaPM2019}.

The review of the literature up to this stage sets us up to introduce the novel concept of significance bands. We will show that the LM principle provides a much simpler approach to think about the absence of a response and to present the information in a convenient graphical way. In addition, we also introduce a straightforward wild block-bootstrap procedure that is simple to implement.

By now, there is a sprawling literature that extends LPs into a variety of new areas. Though we cannot cover all these new developments, we will briefly touch on methods to smooth LPs in the context of producing more efficient responses. We will also discuss applications with panel data and the inferential challenges that these data introduce before concluding.

\section{Local projections: an introduction} \label{s:intro}
There are many situations in the study of dynamic systems where the analyst is interested in the following statistic:
\begin{align*}
\mathcal{R}_{sy|x}(h, s_0, \delta) \equiv \mathbb E(y_{t+h}|s_t = s_0 + \delta; \bm x_t) - \mathbb E(y_{t+h}|s_t = s_0, \bm x_t) \quad \text{for } h = 0, 1, \hdots, H
\end{align*}
where $s_t$ is the variable that generates the impulse from $s_t = s_0$ to $s_t = s_0 + \delta$ for some initial value  $s_0$. Denote $y_t$ as the outcome variable of interest, whose response to an impulse in $s_t$ we want to characterize. Finally, the vector of variables $\bm x_t$ contains lags of the outcome, lags of the impulse, and lags of any other exogenous or predetermined variables, including the constant and deterministic time trends.

We will refer to $s_t$ interchangeably as the \emph{treatment}, \emph{intervention}, or \emph{shock}. At this point, we assume that the shift from $s_0$ to $s_0 + \delta$ is exogenously determined. In practice, identification assumptions and methods will be required. However, since the emphasis here is on inference, we will often prefer to keep things simple. The conditional mean function, $\mathbb E (\cdot | \cdot)$ can in principle be nonlinear, a case considered by \cite{AngristK2011} and \citet*{AJK2016}. We will mostly focus on linear settings, but it is important to maintain the distinction for now.

We use the notation $\mathcal{R}_{s \rightarrow y|\bm x}(h, s_0, \delta)$ instead of $\mathcal{R}_{sy|\bm x}(h, s_0, \delta)$ when additional identifying assumptions, such as exogeneity of $s_t$ justify a causal interpretation of the difference in conditional means. The notation $\mathcal{R}_{s \rightarrow y|\bm x}(h, s_0, \delta)$ is meant to convey the direction of the intervention $s$ onto the outcome $y$, conditional on $\bm x$. When it is clear, we will simplify the notation to $\mathcal{R}_{s  y}(h)$ to denote the response of $y$ to an impulse in $s$, $h$ periods in the future.

For example,  when $s_t \in \{0,1\}$ is randomly assigned, the conditional expectation $ E(y_{t+h}|s_t , \bm x_t)$ does not depend on $\bm x_t$ and only takes two values.  Hence, a natural estimate of the response is simply:
\begin{align*} 
\mathcal{R}_{s \rightarrow y|\bm x}(h, s_0, \delta) = \frac{\sum_{t = h}^T y_{t+h} s_t}{\sum_{t = h}^T s_t} - \frac{\sum_{t = h}^T y_{t+h} (1 - s_t)}{\sum_{t = h}^T (1 -s_t)}; \quad h = 0, 1, \hdots, H.
\end{align*}
\newline
\newline

For $h = 0$, one can think of this difference in means as a measure of the average treatment effect in a randomized controlled trial or RCT. The same statistic can be estimated in regression form:
\begin{align} \label{e:lpregsimple}
y_{t+h} = \mu_h + \beta_h s_t + v_{t+h},
\end{align}
where $\mu_h = \mathbb{E}(y_{t+h}|s_t = 0)$ and $\beta_h = \mathbb{E}(y_{t+h}|s_t = 1) -\mathbb{E}(y_{t+h}|s_t = 0) = \mathcal{R}_{s  y}(h)$. 

\autoref{e:lpregsimple} can be easily generalized to allow for $s_t \in \mathbb{R}$ and to include a vector of controls $\bm x_t$ which, even if $s_t$ is randomly assigned, will improve efficiency. Using a linear framework, we arrive at the typical expression of a local projection:
\begin{align} \label{e:lp}
y_{t+h} = \mu_h + \beta_h s_t + \bm \gamma_h \bm x_t + v_{t+h}.
\end{align}
If $s_t$ is not purely randomly assigned but depends on $\bm x_t$, then including $\bm x_t$ may reduce bias while the effect on the standard errors is ambiguous. The reason is that $\bm x_t$ could be correlated with $s_t$, but have no direct effect on $y_{t+h}$. In that case, including $\bm x_t$ in the regression reduces the variance of $s_t$ without reducing the variance of the regression residual. However, if one is willing to assume that the assignment of $s_t$ is as good as random only given $\bm x_t$, then one can appeal to the conditional independence assumptions carefully spelled out in, e.g., \cite{AngristK2011} and \cite{ AJK2016}. In this scenario, $\bm x_t$ needs to be included in the regression.

If $s_t$ is not randomly assigned even conditional on $\bm x_t$, but there is an instrument or instruments for it, call them $\bm z_t$, then clearly \autoref{e:lp} can be estimated using instrumental variables and we will be more specific later on about the relevance and exogeneity assumptions needed in this case. 

Another obvious extension is to allow for the relationship between $y_{t+h}$ and the right-hand side variables of this expression to be possibly nonlinear, in which case the values of $s_0$ and $\bm x_t$ will be determinative of the actual response estimated. Hence, consider the following general additive conditional mean specification:
\begin{align} \label{e:nl}
y_{t+h} = \mu_h(s_t, \bm x_t; s_0, \delta; \bm \theta_h) + v_{t+h}; \quad h=0,1,\hdots, H\,,
\end{align}
where $s_0$ is the initial state for $s_t$, $\delta$ is the impulse from that initial state, and $\bm \theta$ is the parameter vector. All other components have been previously defined.

Based on \autoref{e:nl}, then:
\begin{align*} 
\mathcal{R}_{sy}(s_0, \delta, \bm x_t, h) = \mu_h(s_t = s_0 + \delta, \bm x_t; \bm \theta_h) - \mu_h(s_t = s_0, \bm x_t; \bm \theta_h); \quad h = 0, 1, \hdots, H,
\end{align*}
where the notation makes explicit the dependence of the response on $s_0$, $\delta$, and $\bm x_t$. For example, consider a simple nonlinear LP (though linear in parameters):
\begin{align} \label{e:lpnleg}
y_{t+h} = \alpha_h + \beta_h s_t + \gamma_h x_t + \phi_h s_t^2 x_t + v_{t+h}
\end{align}
then clearly $\mu_h(s_t = s_0 + \delta, x_t; \bm \theta_h) = \alpha_h + \beta_h (s_0 + \delta) + \gamma_h x_t + \phi_h (s_0 + \delta)^2 x_t$ whereas $\mu_h(s_t = s_0, x_t; \bm \theta_h) = \alpha_h + \beta_h s_0 + \gamma_h x_t + \phi s_0^2 x_t$ and hence $\mathcal{R}_{sy}(s_0, \delta, x_t, h) = \beta_h \delta + \phi_h(\delta^2 + 2\delta s_0)x_t$, which will depend on $s_0$, the initial value of the intervention variable, $\delta$ the size of the intervention, and $x_t$, the value of the current state variable. This example highlights the importance of being careful in how nonlinear responses are calculated as it has direct bearing on how inference needs to be constructed.

Assuming instruments are available and they meet relevance and exogeneity conditions that we will make more precise below, then \autoref{e:nl} can be estimated by the Generalized Method of Moments (GMM) based on the moment conditions:
\begin{align} \label{e:gmm}
\mathbb{E} \left[\left(y_{t+h} - \mu_h(s_t, \bm x_t; \bm \theta_h)\right)\bm z_t \right] = 0; \quad h = 0, 1, \hdots, H.
\end{align}
Furthermore, as we shall explain below, \autoref{e:gmm} for each $h$ can be stacked and estimated jointly as a system. Stacking as a system will prove useful when estimates of the covariance matrix of responses across all horizons is required for simultaneous inference or when it is required for nonlinear models such as the one in \autoref{e:lpnleg}.
\subsection{Properties of the residual}
Inference based on \autoref{e:lpregsimple} obviously depends on the properties of the residual, $v_{t+h}$. We present the main ideas of what differentiates local projections from traditional regression results in as simple a setup as possible. Hence suppose the data are generated by the following covariance-stationary AR(1) model. Using the companion form, the AR(1) encompasses more general AR(p) models and of course, in vector form, general VAR(p) models so the example, while simple, is helpful in thinking of more complex settings:
\begin{align} \label{e:ar1}
y_t = m + \rho y_{t-1} + u_t; \qquad |\rho|< 1; 
\end{align}
where $u_t$ is a zero-mean white noise process with $\mathbb{E} u_t^2 = \sigma^2_u$. All of these assumptions can be relaxed and made much more general but they suffice to make the point.

The goal is to calculate the response of $y_{t+h}$ to a shock $u_t$. Iterating forward $h$ periods into the future on the previous expression, we arrive at a common expression for a local projection, along the lines of that in \autoref{e:lp} where the shock of interest is now $u_t$ instead of some other exogenous variable. Hence we have:
\begin{align} \label{e:lpar1}
y_{t+h} =  \mu_h +  \beta_{h} y_{t} + v_{t+h}, 
\end{align}
where $\mu_h = m (1 + \rho + \hdots + \rho^{h-1})$; $\beta_{h} = \rho^{h}$, and $v_{t+h} = u_{t+h} + \rho u_{t+h-1} + \hdots + \rho^{h-1} u_{t+1}$. Given our assumptions, \autoref{e:lpar1} can be estimated by ordinary least-squares (OLS) and thus, a consistent estimate of $\mathcal{R}_{uy}(h)$ is $\hat \beta_h$. 

However, notice that the residuals have an MA(h) structure that affects the computation of the standard error $\hat \beta_h$ since:
\begin{align*}
\hat \beta_h = \beta_h + \frac{\frac{1}{T-h}\sum_{t=2}^{T-h} y_t v_{t+h}}{\frac{1}{T-h}\sum_{t=2}^{T-h} y_t^2}
\end{align*}
from where
\begin{align*}
(T-h)^{1/2}(\hat \beta_h - \beta_h) = \frac{\frac{1}{(T-h)^{1/2}}\sum_{t=2}^{T-h} y_t v_{t+h}}{\frac{1}{T-h}\sum_{t=2}^{T-h} y_t^2}.
\end{align*}
Given our assumptions on $u_t$  it is easy to see that $\frac{1}{T-h}\sum_{t=2}^{T-h} y_t^2 \overset{p}{\to} \frac{1}{(1 - \rho^2)} \sigma^2_u$, whereas the numerator will converge in distribution to a Normal random variable whose variance, $\omega^2$ is:
\begin{align*}
\omega^2 = Var\left( \frac{1}{(T-h)^{1/2}}\sum_{t=2}^{T-h} y_t v_{t+h}\right) & \approx \sum_{j=-\infty}^\infty \mathbb{E}(y_t v_{t+h} y_{t-j}v_{t+h-j}), 
\end{align*}
where $v_{t+h} = u_{t+h} + \rho u_{t+h-1} + \hdots + \rho^{h-1} u_{t+1}$ and $y_t = u_t + \rho u_{t-1} + \hdots $. Putting the pieces back together we arrive at:
\begin{align} \label{e:asymse}
(T-h)^{1/2} \frac{\sigma^2_u}{\omega(1-\rho^2)}(\hat \beta_h - \beta_h) \overset{d}{\to} N(0, 1).
\end{align}
These derivations show that approximating $\omega$ in finite samples will require a heteroscedasticity and autocorrelation consistent estimator. \cite{Jorda2005} originally proposed using the Newey-West estimator as a simple solution. Since then, several developments that we discuss next have provided more elegant solutions.

\section{LP Inference: The issues}

In thinking about inference, it will be important to clearly outline the objective of the inferential procedure to design the best approach. In this regard at least three obvious objectives come to mind: 
\begin{enumerate}
\item \textbf{Pointwise inference:} How should one assess the precision of individual estimated  response coefficients and the value that they attain? 
\item \textbf{Simultaneous inference:} How should one assess subsets of response coefficients or the trajectory of the response as a whole? 
\item \textbf{Significance:} What is the best way to test the null that the intervention has no effect on the outcome?
\end{enumerate}

Layered on top of these three possible objectives, the analyst should consider the properties of alternative procedures available. \cite{PMWolf2021} and \cite{Xu2023} show that VARs and LPs are asymptotically equivalent when the data are generated by a VAR($\infty$) as long as the lag length is allowed to grow to infinity with the sample size. In finite samples and under the same VAR($\infty$) assumption, \cite*{JST2024} show that, whereas the truncation lag used to estimate the VAR, say $p$, ensures consistency of the first $p$ lags of the VAR, it does not ensure consistency of the impulse response beyond the $p^{th}$ horizon whereas this is not the case for LPs: LPs remains consistent. Moreover, recent work by \cite{PMMOQW2024} shows that  LPs are even robust to misspecification of the truncation lag.

Further, the bias-efficiency trade-off between VARs and LPs, determine the probability coverage properties of each approach. Going back to \cite*{PMMOQW2024}, these authors show that LP inference is robust to relatively large amounts of misspecification even when compared to VARs that are only mildly misspecified. 

How did they arrive at this conclusion? \cite{PMMOQW2024} use a setting where they assume that the data are generated by a VAR where the residuals follow a moving-average structure that vanishes as the sample grows. In large samples, the VAR is clearly correctly specified. However, VARs undercover even when misspecification is so small, that it would be difficult or impossible to detect in small samples. On the other hand, they show that LPs have the correct coverage even when misspecification is so large that it can be easily detected in a finite sample.

Finally, a common concern is not just to derive procedures that are valid in large samples, but to construct methods that account for small sample properties of the data that may be far from the large sample ideal. These will usually include simulation-based inference, such as the bootstrap, as we shall discuss. Bayesian inference is also possible. Although LPs are not a generative model with which to construct the likelihood, \cite{Tanaka2020} and \cite{MirandaRicco2023}, for example, provide Bayesian estimation methods of LPs. LPs can also be used to shrink large dimensional Bayesian VARs toward the responses generated with LPs as \cite{AgrippinoRicco2017} show.

\section{Pointwise Inference} \label{s:inference}

\Cref{s:intro} and \autoref{e:asymse} more specifically, provide intuition for why, generally speaking, the residuals of the LP will have a moving average structure and how this might affect inference as a result. \cite{Jorda2005} proposed using a heteroscedasticity and autocorrelation consistent (HAC) estimator such as \cite{NeweyWest1987}. Much of the literature appears to follow a similar strategy, even when it comes to panel data, where the Driscoll-Kraay  \citep[][]{DriscollKraay1998} covariance estimator is used instead.\footnote{The Driscoll-Kraay estimator can be seen as the extension to panel data of the Newey-West estimator.} However, whether the Driscoll-Kraay estimator is the right approach depends on the dimensions of the panel. Implicitly, the assumption is that $T >> N$, where $T$ is the time dimension of the panel and $N$ is the cross-section dimension. We reserve a more thorough discussion of panel data inference to a later section. Instead, we begin by discussing the recent LP inferential procedures introduced for time series data in the literature.

\subsection{LP-FGLS}

\cite{Lusompa2023} proposes a parametric feasible GLS procedure that directly accounts for the specific MA structure of the residuals. This LP-FGLS procedure can be best explained with the simple AR(1) example in \autoref{e:ar1}. The idea is to use the residuals from the first local projection, $\hat u_t$ as follows.

\paragraph{LP-GLS algorithm}
\begin{itemize}
\item For $h = 1$ estimate:
\begin{align*}
y_t = m_1 + \beta_1 y_{t-1} + u_t \quad \rightarrow \quad \hat \beta_1, \, \hat u_t
\end{align*}
\item For $h = 2$, construct $\tilde y_{t+1} = y_{t+1} - \hat \beta_1 \hat u_t$ and hence estimate:
\begin{align*}
\tilde y_{t+1} = m_2 + \beta_2 y_t + v_{t+1} \quad \rightarrow \quad \hat \beta_2
\end{align*}
\item For $h = 3$, construct $\tilde y_{t+2} = y_{t+2} - \hat \beta_1 \hat u_{t+1} - \hat \beta_2 \hat u_t$ and hence estimate:
\begin{align*}
\tilde y_{t+2} = m_3 + \beta_3 y_t + v_{t+2} \quad \rightarrow \quad \hat \beta_3
\end{align*}
\item For $h > 3$ sequentially generate $\tilde y_{t+h} = y_{t+h} - \hat \beta_1 \hat u_{t+h-1} - \hat \beta_2 \hat u_{t+h-2} - \hdots - \hat \beta_h \hat u_t$ and regress:
\begin{align*}
\tilde y_{t+h} = m_h + \beta_{h+1} y_t + v_{t+h}
\end{align*}
\end{itemize}
The residuals $\hat v_{t+h}$ will be approximately white noise and hence heteroscedasticity robust inference is sufficient.

In a related paper, \cite{BreitungBrugemann2023} propose a similar approach that consists of using the transformation $\tilde y_{t+h} = y_{t+h} - \hat u_{t+h-1}$ and use $\{\hat u_t, \hdots, \hat u_{t+h-2} \}$ as additional regressors. \cite{BreitungBrugemann2023} show that this correction is as efficient as responses estimated with the correct vector autoregression in finite samples. \cite{Lusompa2023} also provides bootstrap versions of the LP-FGLS procedure, which have good efficiency gains relative to standard procedures.

\begin{figure}[t]
\small
\caption{Comparing Newey-West versus FGLS error bands}
{\centering
\includegraphics[width=0.8\textwidth]{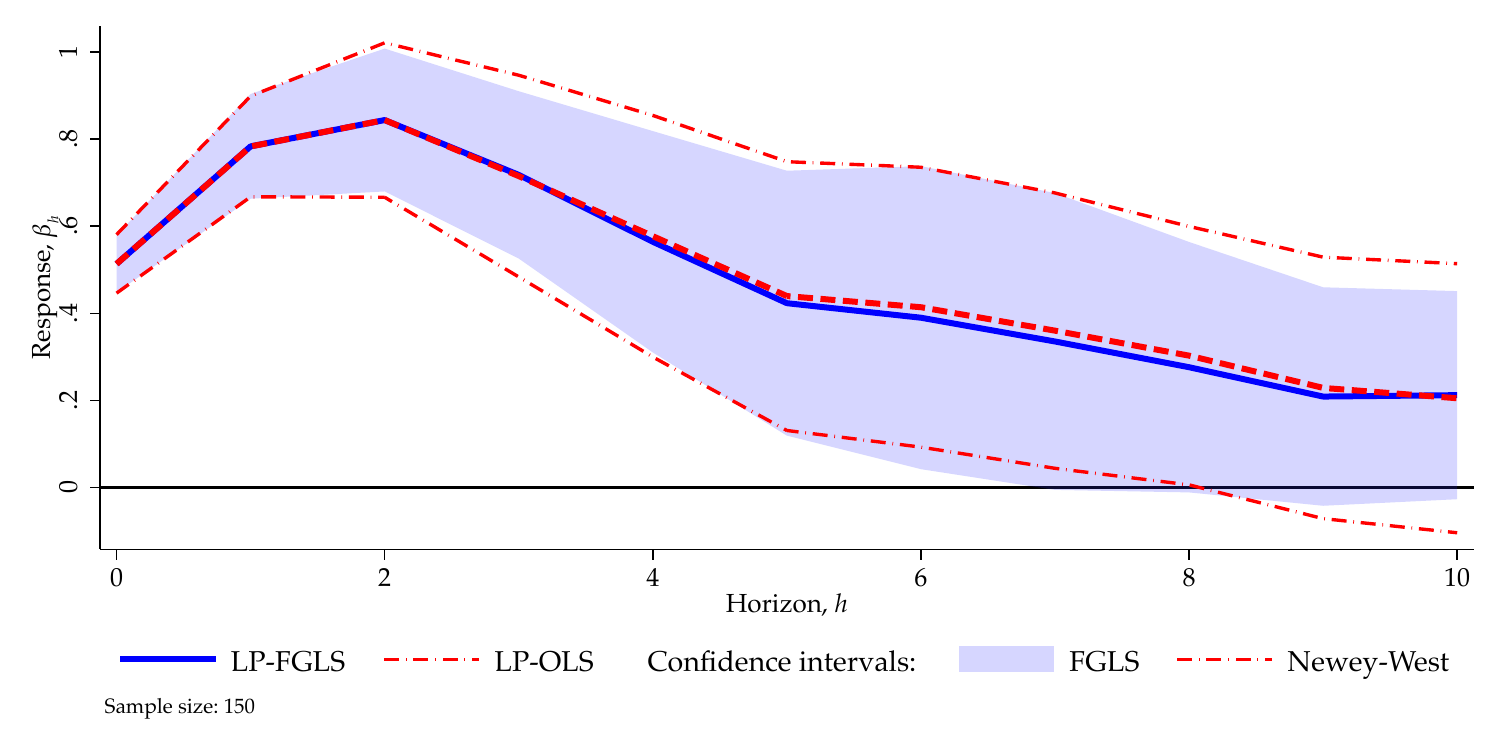} 
\par }
\label{f:fgls}
{\justify\footnotesize
\emph{Notes:} Data generated from a bivariate VAR(1). The simulated sample size is 150 observations after disregarding 1,000 initialization observations. LP-OLS response (dashed line) with Newey-West (dot dashed lines) versus LP-FGLS (solid line) with FGLS (shaded region) error bands at 95\% confidence level. See text.
\par
}
\end{figure}

As an illustration, \autoref{f:fgls} compares estimates of an impulse response using the usual Newey-West correction with estimates based on this LP-GLS procedure. The responses are estimated using simulated data from a simple bivariate model given by:
\begin{align} \label{e:dgpvar1}
\begin{pmatrix}
y_t \\
x_t
\end{pmatrix} = 
\begin{pmatrix}
0.7 & 0.4 \\
0.4 & 0.7
\end{pmatrix}
\begin{pmatrix}
y_{t-1} \\
x_{t-1}
\end{pmatrix} +
\begin{pmatrix}
u^y_t \\
u^x_t
\end{pmatrix}	\,; \qquad u^y_t = e^y_t + e^x_t \,, \quad u^x_t = e^x_t \,;  \quad e^y_t, e^x_t \overset{iid}{\sim} N(0,1) 	\,,
\end{align}
with a sample of 150 observations (after disregarding 1,000 initial observations). \autoref{f:fgls} shows the response of $x$ to a shock to $u^x$ and illustrates that both methods generate almost identical response estimates and very similar error bands with a slight efficiency edge for LP-FGLS. This is expected since the DGP fits quite well with the theoretical background.

\subsection{Lag-augmentation of LPs}  \label{sec:lag_aug}

\cite{MOPM2021} introduce the idea of using lag-augmentation to conduct robust inference with LPs. They show that this procedure is uniformly valid over both stationary and non-stationary data and over a wide range of response horizons. Going back to the stylized model AR(1) in equation \autoref{e:ar1} with $m = 0$, assume $u_t$ is strictly stationary and further assume $E(u_t| \{u_s\}_{s\ne t}) = 0$ \emph{almost surely}. We make these assumptions to follow the setup in \cite{MOPM2021}. Using similar notation to that in their paper, let $\beta(\rho,h)$ denote the LP parameter used to estimate the impulse response $\rho^h$, that is
\begin{align} \label{eq:ar1lp}
y_{t+h} = \beta(\rho,h)\,y_t + \xi_t(\rho, h); \qquad \xi_t(\rho, h) \equiv \sum_{l=1}^h \rho^{h-l} \, u_{t+l}\,.
\end{align} 

Next, \cite{MOPM2021} suggest adding $y_{t-1}$ as an additional regressor to \autoref{eq:ar1lp}.  The purpose of this \emph{lag augmentation} is to make the effective regressor of interest stationary even if the data $y_t$ has a unit root. \cite{MOPM2021} show that, rearranging, the lag-augmented local projection can be written as
\begin{align} \label{eq:lalp}
 y_{t+h} = \beta(\rho, h) \, u_t + \beta(\rho, h+1) y_{t-1} + \xi_t(\rho, h) \,.
\end{align}
Although $u_t$ is stationary and therefore would sidestep distortions to the normal distribution caused by near-to-unity asymptotics, it is not directly observed. However, due to the linear relationship between $y_t$ and $u_t$, the feasible local projection onto $(y_t, y_{t-1})$ provides an estimate of $\beta(\rho, h)$ precisely equal to the one that would be obtained from the projection onto $(u_t, y_{t-1})$. 

Thus, the actual regression to be estimated is:
\begin{align} \label{eq:lalp1}
y_{t+h} = \beta(h) y_t + \beta(h+1) y_{t-1} + \xi_t(h) \quad \to \quad \hat \beta(h), \hat \xi_t(h)\,.
\end{align}

Lag-augmentation has two benefits. As \cite{MOPM2021} show, the distribution of $\hat \beta(h)$ of this feasible lag-augmented local projection is uniformly normal in $\rho \in [-1,1]$ using similar arguments to those used of lag-augmentation in AR inference \citep[see, e.g.,][]{SimsStockWatson1990, TodaYamamoto1995, DoladoLutkepohl1996, InoueKilian2002, InoueKilian2020}. The second benefit is that it simplifies the computation of standard errors (though now the convergence rate will be $T^{1/2}$ instead of $T$).

In particular, it is sufficient to use a heteroskedasticity-robust routine to estimate standard errors for $\hat \beta(h)$, like the usual White correction (in STATA, {\tt{reg}} with the option {\tt{robust}} or even better, {\tt{hc3}}). How can we magically dispense with the moving average structure of the residuals evident in \autoref{eq:ar1lp}? From \autoref{eq:lalp}, note that $u_t$ was assumed to be uncorrelated with past and future values of itself, and therefore the regression \emph{score} $\xi_t(\rho, h) u_t$ is serially uncorrelated. To see this, note that the standard error formula in the ideal regression of \autoref{eq:lalp} would be
\begin{align*}
\hat{\mathbf{s}}_h = \frac{(\sum_{t=1}^{T-h} \hat \xi_t (\rho, h)^2 \hat u^2_t)^{1/2}}{\sum_{t=1}^{T-h} \hat u_t^2} \,.
\end{align*}
But by similar linearity arguments used to justify the feasible augmented local projection, it can be calculated directly from \autoref{eq:lalp1} using White corrected standard errors as indicated.

Several remarks deserve mention. First, \cite{MOPM2021} show that lag-augmented LP inference is relatively robust to persistent data and provides appropriate coverage even at relatively long horizons (as long as $h_T/T \to 0$). Second, lag-augmentation is shown to work more generally when the DGP is assumed to be a VAR($p$) or a vector error correction model (VECM), though we are not aware that similar results have been derived for panel data in settings where the time dimension is larger than the cross section dimension, i.e., $T \gg N$. Of course, when $N \gg T$, asymptotic results are driven by the cross-sectional dimension of the panel and then the asymptotic distribution is normal even when the data are persistent. Third, lag-augmentation can also be applied to identified LPs \citep{PMWolf2021, MOPM2021}.

\begin{figure}[t]
\small
\caption{Comparing Newey-West versus lag-augmented error bands}
{\centering
\includegraphics[width=0.8\textwidth]{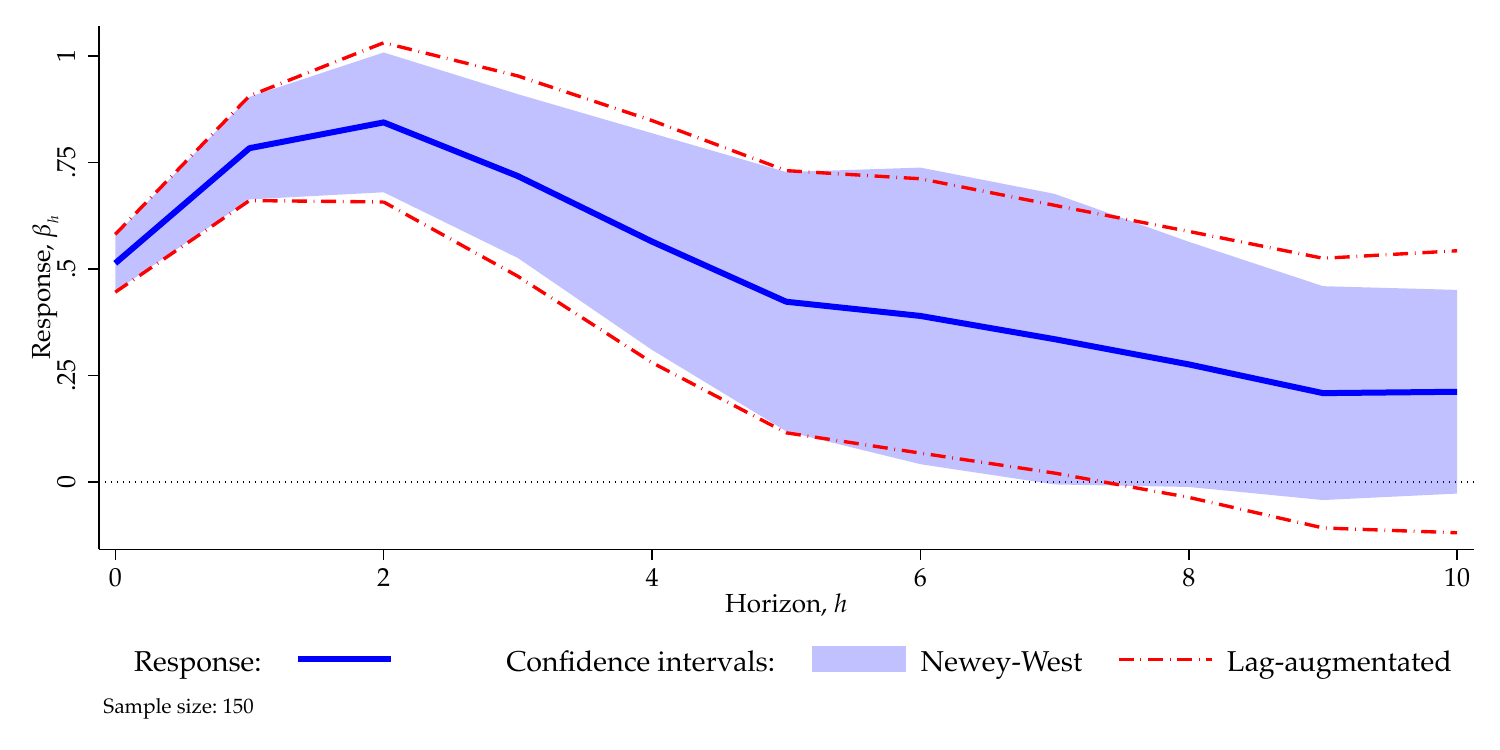} 
\par }
\label{f:nwla}
{\justify\footnotesize
\emph{Notes:} Data generated from a bivariate VAR(1). The simulated sample size is 150 observations after disregarding 1,000 initialization observations. Response shown with Newey-West (shaded region) versus lag-augmented (dot-dashed line) error bands at 95\% confidence level. See text.
\par
}
\end{figure}

As an illustration of how Newey-West and lag-augmented error bands compare, 
we revert back to data simulated from the process shown in \autoref{e:dgpvar1} but this time use LP-OLS estimates with error bands calculated using Newey-West vs lag-augmentation. This is shown in \autoref{f:nwla}.
The figure shows that both methods generate similar error bands. In fact, several experiments (not reported here) suggest that, for stationary data, the coverage is very similar between methods. Lag-augmented bands tend to be somewhat more conservative the more persistent the data.

Finally, \cite{MOPM2021} provide bootstrap procedures that we briefly sketch here though the reader should go to the original source for details. Suppose that one wants to provide inference for an impulse response estimated with lag-augmented LPs for which one can also obtain the standard error as described earlier (i.e., using White corrected standard errors). \cite{MOPM2021} then suggest estimating the corresponding VAR($p$).\footnote{One can also bias-adjust the VAR coefficients using the correction by \cite{Pope1990}.} This VAR will serve two purposes. One is to construct the equivalent response to that estimated with LPs, whose difference is then used to construct the $t$-ratio using the LP standard error. The second is to generate bootstrap replicates of the data using a parametric wild bootstrap \citep[see, e.g.,][]{GoncalvesKilian2004} based on the VAR($p$). Using these bootstrap replicates, then one estimates the lag-augmented LP responses and their standard errors. These are the ingredients necessary to then construct a percentile-$t$ confidence interval as usual. 

\section{Joint inference}

Error bands, as they are overwhelmingly presented in the literature, are the result of inverting the t-statistics of individual hypotheses, as we have seen. When the interest turns to assessing the overall trajectory of the impulse response, the question becomes one of simultaneous inference. Testing joint hypotheses requires the covariance matrix of the response coefficients. Thus, we begin this section by providing a system Generalized Method of Moments (GMM) setup that will allow us to obtain this covariance matrix. Moreover, it will also allow us to consider instrumental variable estimation methods. 

Although joint hypotheses tests are straightforward to construct (under rather general assumptions), presenting the evidence graphically, as we did with error bands, is not. After introducing the GMM estimator for LPs, we discuss two approaches to achieve this, one based on Scheff\'e's S-method, and another based on a sup-$t$ test. Both approaches generate more conservative error \emph{bounds} designed to accommodate hypotheses tests on subsets of response coefficients. In practice, these methods have found less echo in the literature precisely because the bounds are wider. However, some of the insights from these procedures are valuable.

\subsection{LP-GMM} \label{s:GMM}
GMM provides a convenient framework to estimate LPs jointly, as we advanced when discussing \autoref{e:gmm}. Let $\bm y_t(H) = (y_t, \hdots, y_{t+H})'$ be an $(H+1) \times 1$ vector that collects the outcome variable observed at increasingly distant horizons into the future. Let $S_t = I_{H+1} \otimes s_t$ where $I_{H+1}$ is the identity matrix of order $H+1$ and $s_t$ is the intervention variable. We collect the error terms in $\bm v_t(H) = (v_t, \hdots, v_{t+h})'$. The response coefficients are $\bm \beta = (\beta_0, \hdots, \beta_H)'$. Exogenous and predetermined variables collected in $\bm x_t$ could be easily included by defining $X_t = I_{H+1} \otimes \bm x_t$. However, for simplicity, we will mostly set them aside for the presentation. In linear settings, we could simply invoke the Frisch-Waugh-Lovell theorem and orthogonalize the outcome and the intervention with respect to the controls before proceeding. Finally, suppose an $1 \times l$ vector $\bm z_t$ of instrumental variables is available with $l \ge 1$. Hence we construct $Z_t = I_{H+1} \otimes \bm z_t.$ If no instruments are available, but $s_t$ is exogenously determined (perhaps conditional on controls), then one can simply set $Z_t = S_t$.

Using these definitions, the population moment conditions of the system of local projections can be expressed as:
\begin{align*}
E[Z_t'(\bm y_t(H) - S_t \bm \beta)] = 0.
\end{align*}
Thus, the corresponding finite sample GMM problem can be written as:
\begin{align*}
\min_\beta \left[ \frac{1}{N} \sum_{t=t_0}^{T^*} Z_t'(\bm y_t(H) - S_t \bm \beta) \right]' \hat \Lambda ^{-1} \left[ \frac{1}{N} \sum_{t=t_0}^{T^*} Z_t'(\bm y_t(H) - S_t \bm \beta) \right],
\end{align*}
with $N = T^* - t_0$, where $t_0$ denotes the first observation available after accounting for possible lags in the control set and $T^* = T-H-1$. One could choose to set $\hat \Lambda = I_{l\times (H+1)}$. The estimator, referred to as the equally weighted estimator, yields consistent estimates of $\bm \beta$, but the covariance matrix is not valid for inference, as is well-known. 
\newline
\newline
\newline

The optimal weighting matrix, correcting for heteroscedasticity and autocorrelation with, for example, a Barlett correction, is:
\begin{align*}
\hat \Lambda = \hat \Gamma_0 + \sum_{j=1}^J K(j) (\hat \Gamma_j + \hat \Gamma_j'); \quad K(j) = \left[ 1 - \frac{j}{J+1} \right]; \quad \hat \Gamma_j = \frac{1}{N} \sum_{t_0}^N Z_t'\hat{\bm v}_t(H)\hat{\bm v}_{t-j}'(H) Z_{t-j}.
\end{align*}
where $\tilde{\bm v_t}(H)$ refers to the residuals based on the equally weighted estimator, which we know to be consistent. 

\cite{StockWatson2018} and \cite{PMWolf2022} provide appropriate relevance and exogeneity conditions (under relatively general assumptions for the underlying DGP) for the analysis of LPs. 
In our model, these can be stated as:

\begin{itemize}
\item \textbf{\emph{Relevance:}} $E(\bm z_t's_t| \bm x_t) \neq \bm 0$
\item \textbf{\emph{Exogenenity:}} $E(Z_{t}' \bm v_t(H)|X_t) = \bm 0$
\end{itemize}
where we explicitly include the controls in the conditioning set to make clearer their influence (despite having proceeded by projecting their influence away in a first stage or simply when they are subsumed in the vector of instruments, $\bm z_t$).
Our exogeneity condition is satisfied 
under the contemporaneous and lead-lag exogeneity conditions 
of \cite{StockWatson2018}. 

Based on the relevance and exogeneity assumptions just stated and standard results from the algebra of GMM, estimates of the impulse response can be obtained from:
\begin{align*}
\hat{\mathcal{R}}_{sy} = \hat{\bm \beta} = \left(\frac{1}{N} \sum_{t_0}^{T^*} S_t'Z_t \hat \Lambda^{-1} Z_t'S_t \right)^{-1} \left(\frac{1}{N} \sum_{t_0}^{T^*} S_t'Z_t \hat \Lambda^{-1} Z_t'\bm y_t(H) \right),
\end{align*}
which will be consistent and asymptotically normally distributed with approximate covariance matrix given by:
\begin{align} \label{e:omegaGMM}
\hat \Omega_\beta = \left[ \left(\frac{1}{N} \sum_{t_0}^{T^*} S_t'Z_t \right) \hat \Lambda^{-1} \left(\frac{1}{N} \sum_{t_0}^{T^*} Z_t'S_t \right)\right]^{-1}.
\end{align}

Joint estimation of the impulse response using this GMM set-up is useful as it provides an estimate of the covariance matrix, $\hat \Omega_\beta$, a key ingredient to construct any joint hypothesis test of interest. It will also be an important ingredient in the construction of error bands that accommodate simultaneous inference, as the next section briefly explains. 

\subsection{Simultaneous inference}

Analysts interested in testing features of the impulse response involving more than one parameter can set up the appropriate hypothesis test as usual using the F-test, for example, and report the results of the test in a table. However, how should one represent simultaneous inference graphically if one were interested in representing bounds with appropriate probability coverage that accommodate a variety of hypothesis tests an analyst may conduct? Impulse response coefficients are correlated. Thus, the usual practice of inverting the t-ratio to display error bands will not provide the correct probability coverage. The correct approach requires that we construct error bands that account for the simultaneous nature of the family of hypotheses of interest.

This problem was highlighted by \cite{Jorda2009}. His solution relied on Scheff\'e's multiple comparison approximation or S-method \citep{Scheffe1953}. Asymptotically, the sum of the squares of the t-ratios of the LP is approximately $\chi^2_H$ distributed. Thus, the critical value with which to construct simultaneous error bands is $(c_\alpha^2/H)^{1/2}$ where $c_\alpha^2$ refers to the critical value of a $\chi^2_H$. The advantage of this method is that it does not require simulation methods to obtain the critical value.

More recently, \cite{MontielOleaPM2019} proposed a more efficient approximation based on the sup-$t$ procedure. Although this method requires simulation methods, it is shown to produce the narrowest bands for a given probability coverage. In particular, let $\bm \beta = (\beta_1, \hdots, \beta_H),$ and assume $\bm \beta \overset{d}{\to} N(0, \Omega_\beta).$ The goal is to find $c_\alpha$ such that the error bands defined by:
\begin{align*}
\hat{\mathcal{B}}(c_\alpha) = [\hat \beta_0 - c_\alpha \hat \Omega_{0,0}, \hat \beta_0 + c_\alpha  \hat \Omega_{0,0}] \times \hdots \times [\hat \beta_H - c_\alpha  \hat \Omega_{H,H}, \hat \beta_H + c_\alpha  \hat \Omega_{H,H}] 
\end{align*}
such that:
\begin{align*}
P(\bm \beta \in \hat{\mathcal{B}}(c)) \ge 1 - \alpha
\end{align*}
\cite{MontielOleaPM2019} show that:
\begin{align*}
P(\bm \beta \in \hat{\mathcal{B}}(c)) \to P \left( \max_{h=0,1, \hdots, H} \left| \Omega_{h,h}^{-1/2} V_h \right| \le c_\alpha \right); \quad \bm V = (V_0,V_1, \hdots, V_H)' \sim N(0, \Omega_\beta).
\end{align*}
Although the distribution of $ \max_{h=0,1, \hdots, H} \left| \Omega_{h,h}^{-1} V_h \right|$ is unknown, it is relatively easy to obtain any desired quantile of this distribution. Thus, error bands for simultaneous inference can be constructed according to the following algorithm:
\newline
\newline

\noindent \textbf{Plug-in sup-$t$ algorithm}
\begin{itemize}
\item \textbf{Step 1:} Draw $M$ i.i.d. normal vectors $\hat V^{(m)} \sim N_H(\bm 0_H, \hat \Omega_\beta), \, m = 1, \hdots, M.$
\item \textbf{Step 2:} Define $\hat q_{1-\alpha},$ the empirical $1 - \alpha$ quantile of $ \max_{h=0,1, \hdots, H} \left| \Omega_{h,h}^{-1/2} V_h \right|$ across $m = 1, \hdots, M.$
\item \textbf{Step 3:} Construct the error bands for each $h$ as: $[\hat \beta_h - \hat q_{1-\alpha} \hat \Omega_{h,h}, \hat \beta_h + \hat q_{1-\alpha} \hat \Omega_{h,h}]; \, h = 0,1, \hdots, H$
\end{itemize}
\cite{MontielOleaPM2019} further provide bootstrap and Bayesian versions of this procedure. The interested reader should consult their paper for more details. 

\begin{figure}[t]
\small
\caption{Comparing Newey-West versus sup-$t$ error bands}
{\centering
\includegraphics[width=0.8\textwidth]{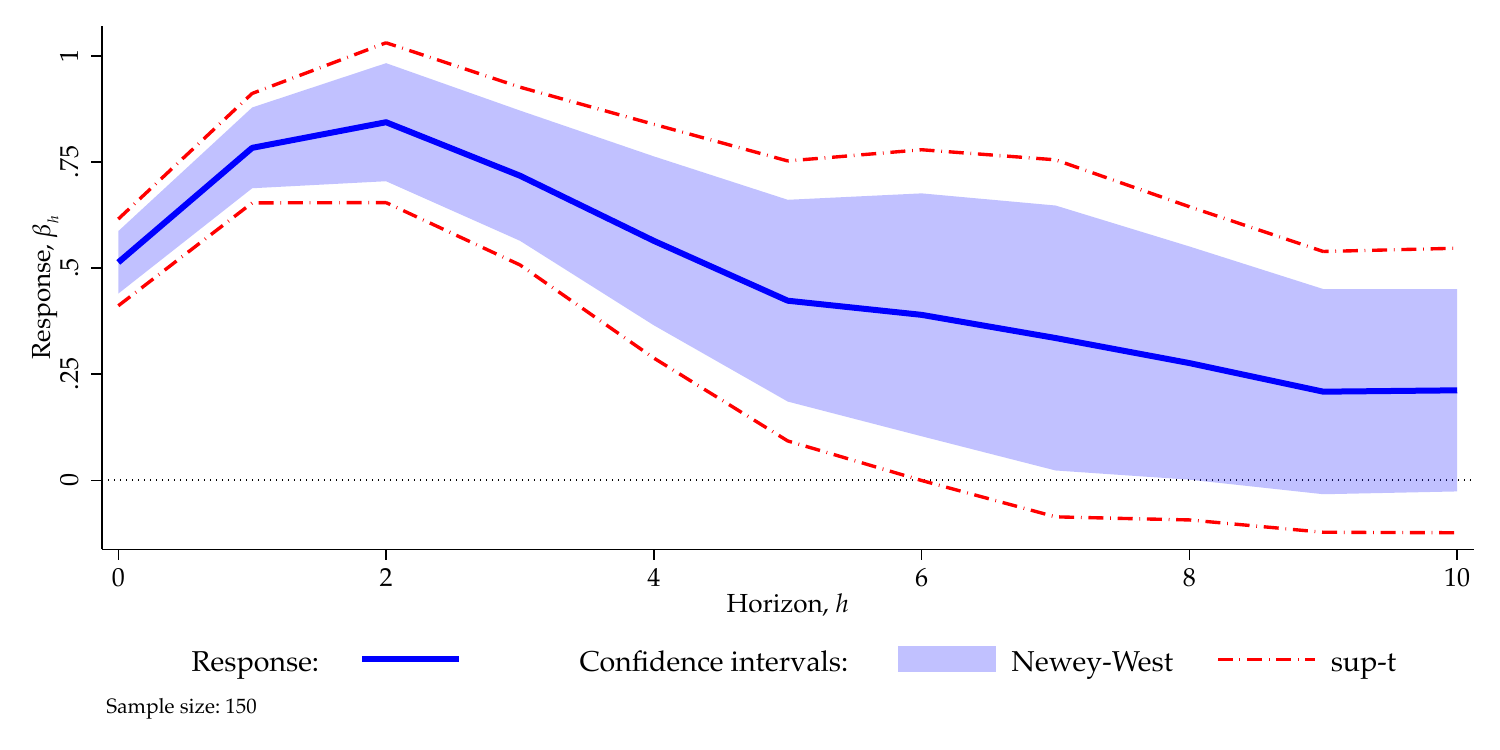} 
\par }
\label{f:nwsim}
{\justify\footnotesize
\emph{Notes:} Data generated from a bivariate VAR(1). The simulated sample size is 150 observations after disregarding 1,000 initialization observations. Response shown with Newey-West (shaded region) versus sup-$t$ (dashed-dotted lines) error bands at 95\% confidence level. See text.
\par
}
\end{figure}

To get a sense of how much wider the sup-$t$ bands are, we revisit the example of \autoref{f:fgls} and \autoref{f:nwla}. We simulate data from the same model and then construct 95\% sup-$t$ bands using joint estimation of all responses simultaneously to obtain the resulting covariance matrix. In \autoref{f:nwsim}, the sup-$t$ bands are shown as dashed blue lines. Relative to the usual Newey-West bands, shown as a shaded region in red, it is easy to see that they are wider but not by a very large amount.

\section{Significance}

In a randomized controlled trial (RCT), a common hypothesis of interest is to test the null that the treatment is ineffective. Similarly, an impulse response can be thought of as the response to a treatment observed over time and therefore, a common hypothesis of interest will be to assess whether the response is different from zero, just as in the RCT example. It turns out that under the null, the distribution of the appropriate hypothesis test can simplify the construction of \emph{significance} bands considerably.

To understand the basic issues, we use a simple example where the controls are set aside to keep the notation simple. For example, we may presume that the outcome $y_{t+h}$, the intervention, $s_t$, and the instrument $z_t$ have been previously orthogonalized with respect to the controls $\bm x_t$ based on the Frisch-Waugh-Lovell theorem. Thus, consider the following local projections estimated by instrumental variables regression:
\begin{align*} 
y_{t+h} = s_t \beta_h + v_{t+h} \quad \text{for } h = 0, 1, \hdots, H - 1; \quad t=1, \hdots, T
\end{align*}

\newpage
We assume that $z_t$, meets the conditions for relevance and exogeneity, as well as an exclusion restriction. Then we postulate that:

\begin{itemize}
\item \textbf{\emph{Relevance:}} $E(s_t \, z_t) \ne 0$.
\item \textbf{\emph{Exogeneity:}} $E(v_{t+h} \, z_{t}) = 0 \,\,\, \forall h=0, 1, \hdots, H - 1$.
\end{itemize}

Depending on the setting, $z_t$ may include $s_t$ itself, such as when $s_t$ is an observable shock, and then the discussion returns to a more traditional OLS setting. Or if $s_t$ is, conditional on $x_t$, sequentially exogenous. This would be the case in a recursive identification scheme. We further assume that $y_t$, $s_t$, and $z_t$ are covariance stationary. This assumption is not necessary to ensure consistency of the local projection, but will make deriving our inferential procedures and the presentation in this section straightforward.

Based on this simple set up, the instrumental variable estimator for $\beta_h$ can be written as:
\begin{align} \label{e:betah}
\sqrt{T-h} (\hat \beta_h - \beta_h) = \frac{(T-h)^{-1/2} \sum_1^n z_t \> y_{t+h}}{(T-h)^{-1} \sum_1^n z_t \> s_t} \quad \text{for } h = 0,1,\hdots, H-1,
\end{align}
where we note that we will evaluate the statistic under the null $H_0:\beta_h = 0$. Under standard regularity conditions, and the instrumental variable assumptions for local projections, it is easy to see that:
\begin{align*} 
\frac{1}{T-h} \sum_1^n z_t \> s_t \overset{p}{\to} E(z_t \> s_t) \equiv \gamma_{zs}
\end{align*}
Next, consider the numerator in \autoref{e:betah} evaluated at the null $H_0: \beta_h = 0$:
\begin{align*} 
\frac{1}{(T-h)^{1/2}} \sum_1^n z_t \> y_{t+h} \overset{d}{\to} N(0, s_{\eta,h}^2)
\end{align*}
where $s_{\eta,h}^2$ is given by
\begin{align*} 
s_{\eta,h}^{2}= \lim_{T\rightarrow\infty}Var\left( \frac{1}{(T-h)^{1/2}} \sum_1^n z_t \> y_{t+h} \right) & \> = \sum_{j=-\infty}^\infty E(z_t \> v_{t+h} \> z_{t-j} \> v_{t+h - j})  \notag \\
\end{align*}
and where the RHS is the typical HAC type variance formula. The equality follows from the null hypothesis that $\beta_h = 0$ for $h = 0,1, \hdots, H-1$.  It then follows that the limiting distribution of $\hat\beta_h$ is 
given by 
\begin{align} \label{e:betah_limgen}
\sqrt{T-h} (\hat \beta_h - 0) \overset{d}{\to} N(0, \sigma_h^2); \quad \sigma_h^2 = \frac{s_{\eta,h}^2}{\gamma_{zs}^2}; \quad \forall h 
\end{align}
From \autoref{e:betah_limgen} it is easy to derive a $1-\alpha$ percent band around the zero null so that:
\begin{align*}
P\left[\zeta_{\alpha/2}\,\frac{\sigma_h}{\sqrt{T-h}} < \hat \beta_h  < \zeta_{(1-\alpha/2)}\,\frac{\sigma_h}{\sqrt{T-h}} \right] =1 - \alpha
\end{align*}
where $\zeta_{\alpha/2}$ is the critical value of a standard normal variable at $\alpha/2$ and for a standard normal, $\zeta_{1 - \alpha/2} = -\zeta_{\alpha/2}$, as is well known. 

To construct feasible confidence intervals we need to replace $\sigma_h$ with an estimate. The LM principle requires that $\sigma_h$ be estimated using the conventional formula for HAC robust standard errors for the just identified two-stage least squares estimator, but evaluated at $\beta_h = 0$. This is accomplished by estimating $s_{\eta,h}^2$ with the long-run variance of $\eta_{t,h} = z_t y_{t+h}$. The estimator for $\sigma_h^2$ then is based on 
\begin{align} \label{e:HAC-var}
\sigma_h^2 = \frac{s_{\eta,h}^2}{\gamma_{sz}^2}.
\end{align}

When plotting a significance band of an impulse response up to $H$ periods, we are essentially conducting a joint hypothesis test. Intuitively, the more horizons considered, the more likely it is to spuriously reject the null when the null is true in a finite sample. A simple way to address this issue is with a Bonferroni adjustment as proposed in \cite{Dunn1961} so that the significance bands for each $\hat \beta_h$ become:
\begin{align*}
\left[ \zeta_{\alpha/2(H+1)} \frac{\sigma_h}{\sqrt{T-h}}, \,\, \zeta_{1 -\alpha/2(H+1)} \frac{\sigma_h}{\sqrt{T-h}} \right].
\end{align*}
The joint probability that the estimated impulse response lies within the confidence band is given by:
\begin{align*}
P \left( \bigcap_{h=0}^{H}  \left\{ \zeta_{\alpha/2(H+1)} \,\frac{\sigma_h}{\sqrt{T-h}} < \hat \beta_h  < \zeta_{(1-\alpha/2(H+1))}\,\frac{\sigma_h}{\sqrt{T-h}} \right\} \right) \ge 1 - \alpha
\end{align*} 
where the inequality holds in large samples and when the null hypothesis of a zero response is true. Similarly, the test of the joint hypothesis that all response coefficients are zero rejects when:
\begin{align*}
\hat \beta_h \not\in \left[ \zeta_{\alpha/2(H+1)} \frac{\sigma_h}{\sqrt{T-h}}, \,\, \zeta_{1 -\alpha/2(H+1)} \frac{\sigma_h}{\sqrt{T-h}} \right]
\end{align*}
for at least one $h$. By the same argument, it follows that the size of such a test is not more than $\alpha$ in large samples.

Under stronger assumptions such as independence between $z_t$  and $v_s$ for all $t$ and $s$, or homoskedasticity of $v_{t+h}$ such that  $E(v_{t+h - j} v_{t+h}|z_t, z_{t-j} \> )=E(v_{t+h - j}v_{t+h})$  a further simplification of the expression for  $s_{\eta,h}^2$ is possible. We obtain
\begin{align} \label{e:simpleomega}
s_{\eta}^2=\sum_{j=-\infty}^\infty E(z_t \> v_{t+h} \> z_{t-j} \> v_{t+h - j}) & \> = \sum_{j=-\infty}^\infty E(z_t \> z_{t-j}) E(v_{t+h} \> v_{t+h-j}) \\ 
& \>  =\sum_{j=-\infty}^\infty \gamma_{z,j} \> \gamma_{y,j} \notag 
\end{align}
where the equality follows from the independence between $z_t$ and $v_{t+h}$. We define $\gamma_{z,j}$ and $\gamma_{y,j}$ as the $j^{th}$ autocovariances of $z$ and $y$ respectively. Importantly, note that $\omega$ is no longer a function of the horizon $h$ under these additional restrictions. The implication is that under the additional restrictions of homoskedasticity or independence, the significance bands will be constant as a function of the horizon h. 

Under these stronger conditions we can write \autoref{e:betah_limgen} under the null hypothesis as:
\begin{align} \label{e:betah_f}
\sqrt{T-h} (\hat \beta_h - 0) \overset{d}{\to} N(0, \sigma^2); \quad \sigma^2 = \frac{\sum_{j=-\infty}^\infty \gamma_{z,j} \gamma_{y,j}}{\gamma_{zs}^2} = \frac{s_{\eta}^2}{\gamma_{zs}^2}; \quad \forall h 
\end{align} 
A simple example provides further intuition and a connection to well-known results. In the special case where $z = s$, and $y$ and $s$ are serially uncorrelated,  \autoref{e:simpleomega} simplifies even further to:
\begin{align*}
\sigma^2 = \frac{\gamma_{y,0}}{\gamma_{s,0}}
\end{align*} 
Thus, when $y = s = z$ and $y$ is a white noise and hence $\gamma_{y,0} = \gamma_{s,0}$ so that $\sigma^2 = 1$, the local projection estimator is simply an estimator of the autocorrelation function. Hence, applying the same derivations as in \autoref{e:betah_f}, it is easy to see that one recovers the well known\footnote{Not Barlett corrected.} bands for the autocorrelogram of $y$. Specifically, focus on $h = 1$ in the special case that $y$ is a white noise but one estimates an AR(1) model:
\begin{align} \label{e:ar1dist}
\sqrt{n} (\hat \rho - 0) \overset{d}{\to} N(0, 1). 
\end{align} 
This is the well known case where the 95\% asymptotic significance bands in a correlogram are calculated as $\pm 1.96 \times 1/\sqrt{n}$ and provides a nice window into our proposed procedures. Importantly, notice that the bands do not depend on the horizon (in fact, they also do not depend on the variance in this special case). Whenever an autocorrelation coefficient exceeds the band, the interpretation is that said coefficient can be deemed to be different from zero. This, of course, means that the hypothesis that the impulse/treatment has no effect on the outcome can be rejected. 

\subsection{Practical implementation}

Constructing significance bands in practice based on the results from the previous section is straightforward and can be implemented using standard statistical software. We provide a STATA example to illustrate this point and that corresponds to the figures displayed in the paper. The basic steps can be summarized as follows:
\bigskip

\hrule
\smallskip
\textbf{Significance bands using asymptotic approximations}
\smallskip
\hrule
\begin{enumerate}
\item Calculate the sample average of the product $s_t \, z_t$. Call this $\hat \gamma_{sz}$.
\item Construct the auxiliary variable $\eta_{t,h} = y_{t+h} \,z_t$ and regress $\eta_{t,h}$ on a constant. The Newey-West estimate of the standard error of the intercept coefficient is an estimate of $s_{\hat{\eta},h}$.
\item An estimate of $\sigma/\sqrt{T-h}\,$, call it $\hat s_{\beta_h}\,$, is therefore:
\begin{align*}
\hat s_{{\beta}_h} = \frac{\hat s_{{\hat \eta},h}}{\hat{\gamma}_{sz}}
\end{align*}
\item Construct the significance bands as:
\begin{align*}
 \left[\zeta_{\alpha/2(H+1)} \hat s_{{\beta}_h}, \,\, \zeta_{1 - \alpha/2(H+1)} \hat s_{\beta_h} \right]
\end{align*}
\end{enumerate}
\hrule
\bigskip

A bootstrap procedure is equally easy to construct. Note that we do not take a position on the data generating process (DGP). Therefore, we apply the bootstrap directly to step 2 of the previous construction of the significance band. Because of the time series dependence and the possible existence of heteroscedasticity, we will use a wild-block bootstrap \citep[see, e.g.][]{GoncalvesKilian2007}. The STATA implementation only requires a few lines of code. Thus, the entire procedure can be described as follows:
\bigskip

\hrule
\smallskip
\textbf{Significance bands using the Wild-Block Bootstrap}
\smallskip
\hrule
\begin{enumerate}
\item Calculate the sample average of $s_t \, z_t$. Call this $\hat \gamma_{sz}$.
\item Construct the auxiliary variable $\eta_{t,h} = y_{t+h} \,z_t$ and regress $\eta_{t+h}$ on a constant. The Wild Block bootstrap estimate of the standard error of the intercept coefficient is an estimate of $s_{\hat{\eta},h}$.
\item An estimate of $\sigma/\sqrt{T-h}\,$, call it $\hat s^b_{\beta_h}\,$, is therefore:
\begin{align*}
\hat s^b_{{\beta}_h} = \frac{\hat s^b_{{\hat \eta},h}}{\hat{\gamma}_{sz}}
\end{align*}
\item Construct the significance bands as:
\begin{align*}
\left[\zeta_{\alpha/2(H+1)} \hat s^b_{{\beta}_h}, \,\, \zeta_{1 - \alpha/2(H+1)} \hat s^b_{\beta_h} \right]
\end{align*}
\end{enumerate}
\bigskip

\subsection{Application: the response of shelter inflation to monetary policy}
We showcase these procedures with a simple application to shelter inflation. As the COVID-19 pandemic was winding down, inflation measured by the personal consumption expenditures (PCE) index, excluding food and energy (core PCE inflation), peaked at around 5.5\% on March 2022. In response, the Federal Reserve raised the federal funds rate and subsequently inflation declined to 2.6\% by June 2024 (the last data point as of the writing of this paper). However, inflation has been slow to travel back to the 2\% target, in large part, because shelter inflation (which mainly measures rents that tenants face and owners de facto pay themselves) has been slow to decline.

Hence, we evaluate how responsive shelter inflation is to interest rates. We use the series of monetary shocks recently provided by \cite{BauerSwanson2023}. These shocks are obtained from high frequency financial data after removing information effects. The sample available is January 1988 to December 2019 so as to avoid polluting our estimates with the COVID-19 pandemic. We use as controls 12 lags of shelter Personal Consumption Expenditures (PCE) inflation, the unemployment rate (to control for aggregate conditions in the economy), and lags of the federal funds rate. The left-hand side variable is the long difference of 100 times the log of shelter PCE price index (PCEPI), that is $100(y_{t+h} - y_{t-1})$, where $y_t$ is the log of shelter PCEPI. This means that the response reflects the cumulative percentage change in the level of shelter prices up to $h$ periods since the shock.

\begin{figure}[h!]
\begin{center} 	
	\small
	\caption{Response of PCE shelter inflation in percent to a Bauer-Swanson monetary shock}
	\subcaption*{Significance bands displayed}
	\includegraphics[width=0.8\textwidth]{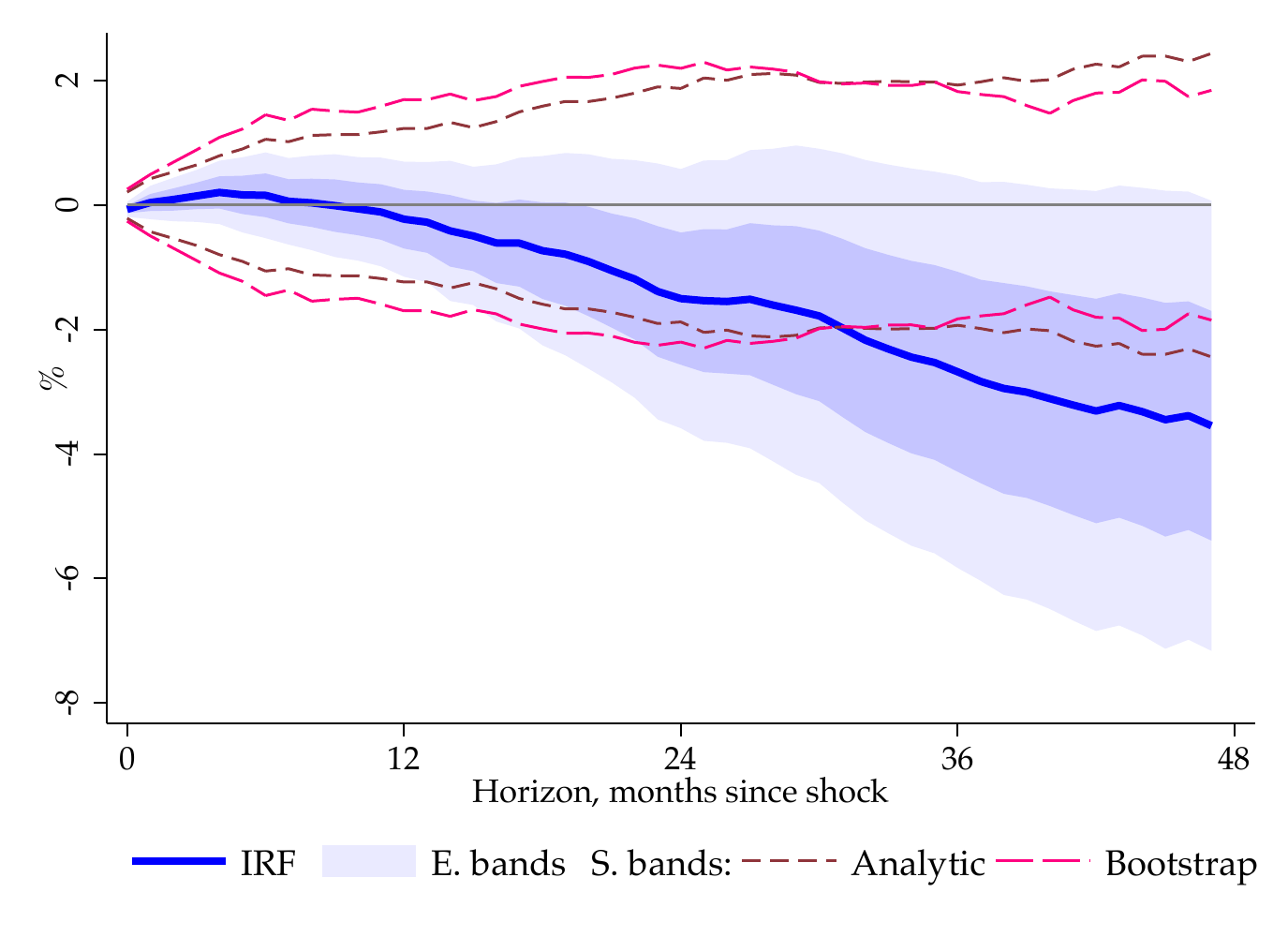} 
	\label{f:inflation}
\end{center}	
	{\justify \scriptsize {\emph{Notes:} cumulative response of shelter PCEPI to a Bauer-Swanson monetary shock \citep{BauerSwanson2023}. The specification includes 12 lags of shelter PCE inflation, the unemployment rate, and the federal funds rate. The sample is January 1998 to December 2019. Traditional, point-wise one and two standard error bands displayed as shaded regions using Newey-West standard errors along with analytic (in maroon, short-dash) and bootstrap (in pink long-dash), Bonferroni adjusted significance bands.  See text.}  
	}
\end{figure}

\autoref{f:inflation} displays estimates of this cumulative response along with Newey-West confidence bands (for 1 and 2 standard deviations in width) along with significance bands estimated with both analytic and bootstrap methods. Based on the conventional error bands displayed in the figure, one might be tempted to conclude that shelter inflation does not respond to monetary shocks. The error bands contain zero throughout the 48 periods displayed.

However, note that the response of shelter PCEPI is almost uniformly negative each month over the 4 years displayed. Indeed, an F-test of the null that the response coefficients are jointly zero is estrenuosly rejected (the p-value is 6.18e-78). And in fact, the significance bands displayed show that, except for about two and one half years after the shock, the response is clearly different from zero. Analytic bands are narrower than bootstrap bands for about 30 periods, after which they are slightly wider.

\subsection{Monte Carlo evidence}
This section presents a couple of simple experiments in graphical form to assess the calculation of significance bands using both the asymptotic approximation and the wild-block bootstrap procedures discussed in the previous section. The data are generated as follows:
\begin{align*}
\begin{cases}
y_t = \beta s_t + 0.75 y_{t-1} + u_{yt} \\
s_t =  0.5 s_{t-1} -0.25 y_{t-1} +  z_t + u_{st} \\
z_t = u_{zt}
\end{cases} \quad u_{yt}, \, u_{st}, \, u_{zt} \, \sim \, N(0, 1); \quad \beta \in \{0, 0.25, 0.50, 0.75 \}
\end{align*}

This simple system encapsulates several features. First, the treatment variable, $s_t$, affects the outcome, $y_t$, contemporaneously. The outcome is itself serially correlated with a coefficient $0.75$. The idea is to have internal propagation dynamics. Next, the intervention responds to feedback from the value of the outcome in the previous period, but also has some internal propagation dynamics. In addition, movements in the intervention are caused by the exogenous variable $z_t$, which will act as our instrumental variable. Finally, the coefficient $\beta$, which captures the effect of the treatment on the outcome, has values between 0 and 0.75. When $\beta = 0$ we have the null model with which to assess the size of the test. Increasing the value of $\beta$ allows us to assess the power of the significance bands.

\begin{figure}[hbt!]
\begin{center} 	
	\small
	\caption{Significance Bands: Monte Carlo exercise}
	\label{f:mc}
	\subcaption{Sample size: 100}
	\includegraphics[width=0.75\textwidth]{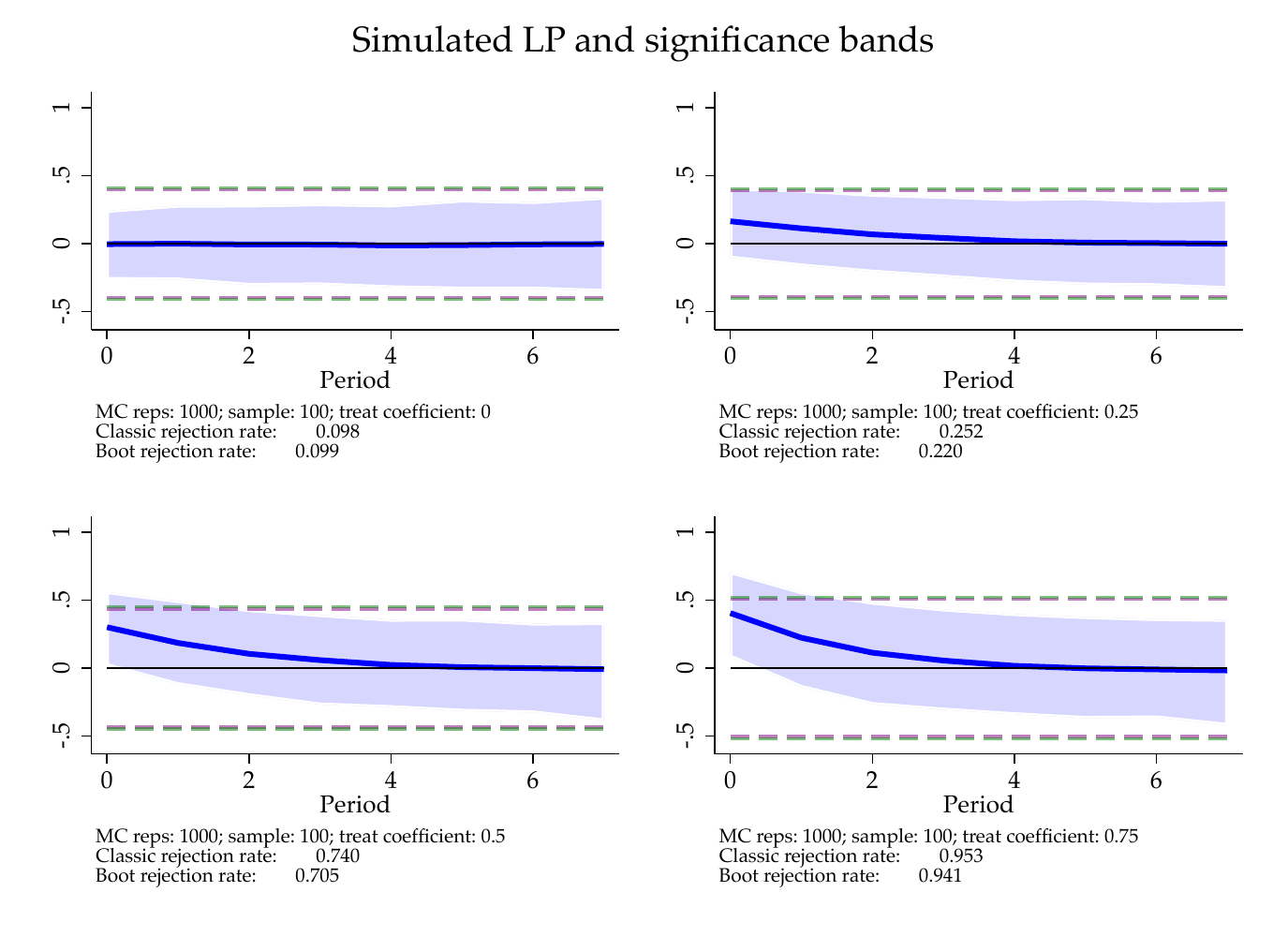} 
	\label{f:mc100}
	\subcaption{Sample size: 500}
	\includegraphics[width=0.75\textwidth]{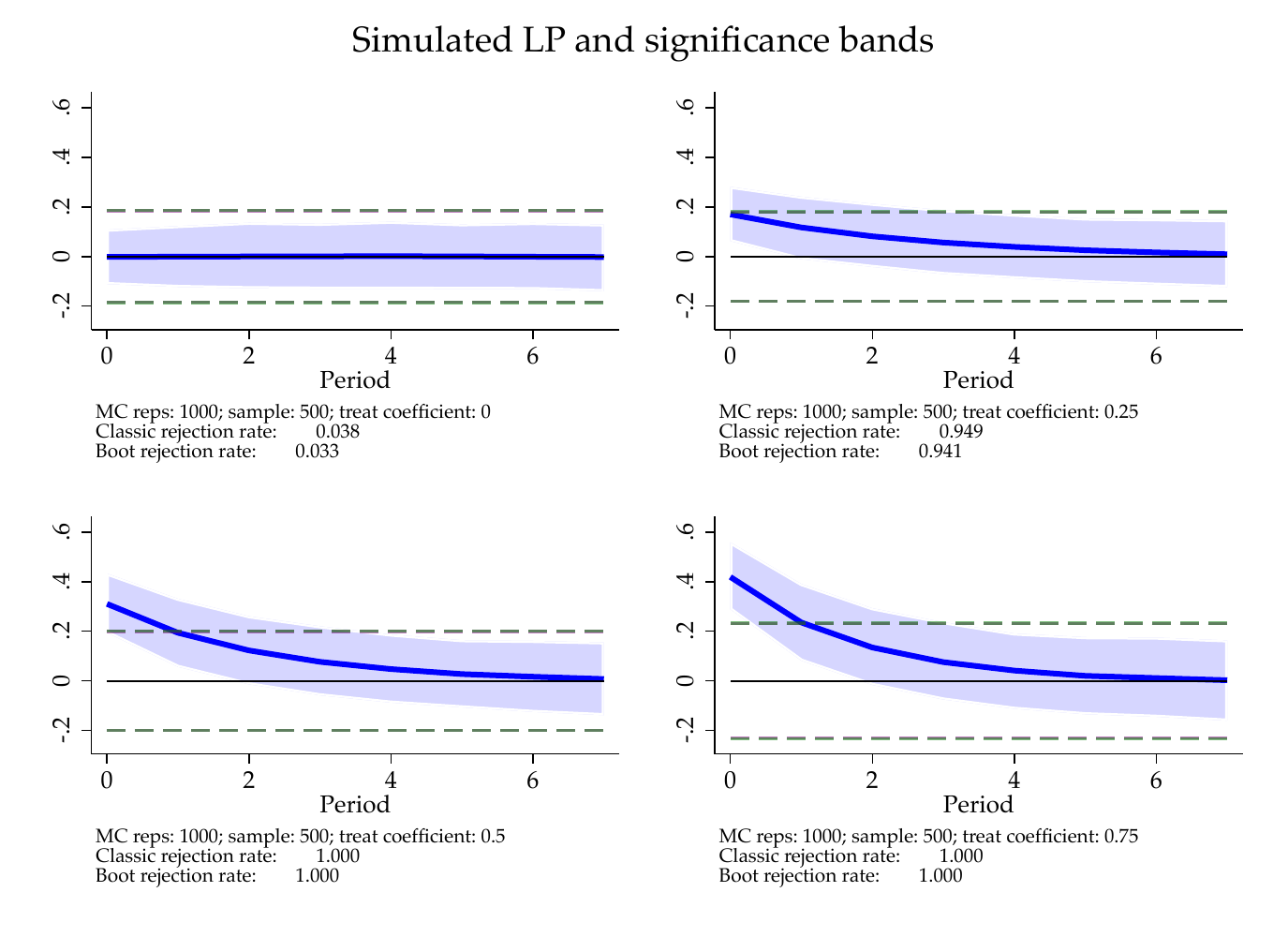}
	\label{f:mc500} 
\end{center}	

	{\justify \scriptsize {\emph{Notes:} Monte Carlo exercise. Sample size = 100 and 500 observations (500 burn-in replications). Significance bands constructed using asymptotic approximations (Classic) and the Wild Block bootstrap (Boot). The rejection rate refers to the share of replications where one or more coefficients exceed the significance bands constructed with each procedure. Treat coefficient refers to the coefficient $\beta$ described in the text. Significance bands constructed at 95\% confidence level. Thus, when Treat coefficient = 0, the rejection rate should be 0.05, otherwise, it should be 1. See text.}  
	}
\end{figure}

We generate samples of 100, and 500 observations with 500 burn-in observations that are discarded to avoid initialization problems. For each sample size and for the different values of $\beta$ we generate 1,000 Monte Carlo replications. The implementation of the wild block-bootstrap is based on 1,000 bootstrap replications as well. For the Newey-West step as well as for the block size in the bootstrap, we use 8 lags. \autoref{f:mc} displays the results for sample sizes of 100 and 500 observations.

\newpage
The figure summarizes quite a bit of information. The shaded bands around the mean estimate of the impulse response showcase the $25^{th}$ and the $975^{th}$ largest values for each coefficient estimate in the Monte Carlo simulation. The dashed lines correspond to the significance bands. Both Newey-West and the bootstrap procedures (using 8 lags) generate nearly indistinguishable values so the differences cannot be seen with the naked eye. For each Monte Carlo exercise, we construct rejection rates for each type of band constructed. The rate is calculated as the share of replications where one or more impulse response coefficients exceed the significance bands.

Several results deserve comment. First, consider the size of the test. We have chosen a rather conservative strategy with a window of size 8 both for Newey-West and for the block-size in the implementation of the bootstrap. As a result, with a small sample of 100 observations, the size is about 10\% instead of the nominal 5\%, though with 500 observations the size is close to 4\%.

However, even with this conservative choice, the power of the test is respectable with a sample size of 100, improving from about 25\% when $\beta = 0.25$ to about 95\% when $\beta = 0.75$. These numbers jump with 500 observations with about 95\% for $\beta = 0.25$ and 100\% even for $\beta = 0.5$.

\section{Other LP settings}
Work on extensions to LPs is ongoing. Two settings in particular, have direct implications for how we think about LP inference. LPs can be seen as a semiparametric estimate of the true impulse response and the reason why they have lower bias. Thus, some authors have proposed methods to either smooth the LPs via some low dimensional series approximation \citep[e.g.][]{BarnichonBrownlees2018, BarnichonMatthes2018}, or by pairing LPs with Bayesian methods to shrink high-dimensional VARs toward the LP response \citep[e.g.][]{Tanaka2020, MirandaRicco2023}.

The other setting that is popular in empirical research is panel data. Because LPs are a single-equation method, they lend themselves well to panel data settings. However, although there is an extensive literature using LPs with panels and there is an extensive literature on inference in panel data settings, little is known about LP inference in panel data settings. In the next sections, we give a brief overview on these less traditional settings.

\subsection{Smoothing local projections}

Smoothing can be used to reduce the uncertainty about the impulse response. Consider the GMM moment condition that we introduced in \autoref{s:GMM}, repeated here for convenience:
\begin{align*}
E[Z_t'(\bm y_t(H) - S_t \bm \beta)] = 0
\end{align*}

\FloatBarrier

\noindent where $\bm \beta$ is a $(H+1) \times 1$ dimensional vector with covariance matrix $\Omega_\beta$, which can be estimated as shown in \autoref{e:omegaGMM}. Smoothing can be thought of as replacing $\bm \beta$ with a lower-dimensional function, say $\phi(h, \bm \theta)$ with $dim(\bm \theta) << dim(\bm \beta)$. In \cite{BarnichonBrownlees2018}, the authors propose using the B-spline method of \cite{EilersMarx1996}, whereas \cite{BarnichonMatthes2018} propose using Gaussian basis functions.  

In general, assuming that one can obtain LP estimates that are asymptotically normal (such as when estimating LPs using the GMM setup in \autoref{s:GMM}) so that $\hat{\bm \beta} \overset{d}{\to} N(\bm \beta, \Omega_\beta)$ and $\hat \Omega_\beta \overset{p}{\to} \Omega_\beta$, then one can setup the minimum distance problem:
\begin{align}
\min_\theta Q(\bm \theta) = \min_\theta (\hat{\bm \beta} - \phi(h; \bm \theta))' \hat \Omega_\beta^{-1}(\hat{\bm \beta} - \phi(h; \bm \theta))
\end{align}
to estimate $\hat \phi(h, \hat{\bm \theta})$ efficiently. Moreover, under standard regularity conditions, the quality of the approximation can be judged with an overidentification restrictions test since $\hat Q(\hat{\bm \theta}) \to \chi^2_q$ with $q = dim(\bm \beta) - dim(\bm \theta)$. The variance of $\bm \theta$ is $\hat \Omega_\theta = (\Phi_0' \hat \Omega_\beta^{-1} \Phi_0)^{-1}$ where $\Phi_0 = \partial \phi(h, \bm \theta)/\partial \bm \theta$ evaluated at the true value $\bm \theta_0$. An example of how smoothing can make response estimates more efficient is \cite{LPMW2022}.

\subsection{Panel data}
Panel data offers more opportunities to explore data using LPs and more opportunities to conduct inference. As we anticipated in \autoref{s:inference}, when the $T \to \infty$ with $N$ fixed or when $N$ grows at a slower rate than $T$, the asymptotic distribution of the LP estimates will be dominated by the time-dimension of the panel and in that case, the counterpart to Newey-West HAC standard errors is to use the \cite{DriscollKraay1998} covariance estimator.\footnote{The command {\tt xtscc} in STATA implements this type of covariance matrix estimation.} 

Cluster robust inference can be used in situations where $N \to \infty$ with $T$ fixed. In such a setting, autocovariances are relatively efficiently estimated and there is no need to specify the residual autocorrelation structure. When $T$ is relatively small, a recommended correction for heteroscedasticity is the wild bootstrap \citep[see, e.g.][]{CameronGM2008, CanaySS2021, MacKinnonOR2019}\footnote{In STATA, this type of boostrap can be implemented with the user supplied command {\tt boottest}.}. If $N$ is relatively large, the asymptotic distribution will be dominated by the cross-sectional dimension of the panel. In that case, stationarity or lack thereof plays no role in computing standard errors.

Relatedly, in a recent paper, \cite{MeiShengShi2023} show that incidental parameter biases \citep{Nickell1981} crop up when the dimensions of the panel $N,T \to \infty$ as $N/T \to c,\, c \in (0, \infty)$. To avoid this bias, these authors suggest using the split panel jacknife estimator of \cite{Dhaene2016} and \cite{Chudik2018}. Denote $\hat \beta_h$ the full sample estimate with fixed-effects and $\hat \beta_h^a$ and $\beta_h^b$ estimates based on splitting the sample along the time series dimension into two halves, $T = T_a + T_b$. Then, the bias corrected estimate of the impulse response is $\tilde \beta_h = 2 \hat \beta_h - 0.5(\hat \beta_h^a + \hat \beta_h^b).$ However, note that a recent paper by \cite{HHKW2024} shows that the split panel jackknife bias correction is generally higher order inefficient and may significantly increase estimator variance in finite samples compared to higher order efficient bias correction methods.

\section{Conclusion}
The error structure of local projections and the kind of hypotheses implicit in the way inference is obtained and communicated requires some care. Residual serial correlation in local projections can be dealt with and more recent developments show that it can be obtained rather easily using lag augmentation and heteroscedasticity robust methods. Bounds for simultaneous inference are also relatively easy to construct, though in most situations we expect that practitioners will simply report formal hypothesis tests.

The significance bands introduced in this review are simple to construct and can be easily displayed alongside the usual confidence bands. While confidence bands inform the reader about the estimation uncertainty of each coefficient, significance bands inform the reader about the significance of the impulse response itself. 

Panel data settings present their own challenges and opportunities. The time series and cross-sectional dimensions of the panel play a critical role in choosing the best inferential procedures. Cluster robust inference generally offers an attractive approach, but cannot always be directly used. Inference in panel data settings is an ever growing field and new developments are constantly arriving to improve existing methods.

Applications and extensions of local projections continue to grow. In this review we are unable to cover every single scenario. However, we hope to have provided the reader with general principles that can then be tailored to each specific extension.

\clearpage
\newpage
\small\singlespacing
\bibliographystyle{authordate1}

\setcounter{secnumdepth}{0}

\renewcommand\bibsection{\section{\refname}}

\appendix

\section{Appendix}

{\small In this appendix we provide some background on the construction
of significance bands proposed in Section 6. }{\small\par}

\subsection{The LM Statistic}

{\small We consider the just identified instrumental variables estimator
discussed in the paper under the constraint that the null $H_{0}:\beta_{h}=0$
holds. The estimator $\hat{\beta}_{h}$ solves the problem 
\[
\underset{\beta_{h}}{\textrm{argmin}}\frac{1}{2}\left(\sum_{t=1}^{T}z_{t}\left(y_{t+h}-s_{t}\beta_{h}\right)\right)^{2}.
\]
 We can set up a Lagrangian to analyze the constrained problem $\beta_{h}=0$
as
\[
L\left(\beta_{h}\right)=\frac{1}{2}\left(\sum_{t=1}^{T}z_{t}\left(y_{t+h}-s_{t}\beta_{h}\right)\right)^{2}+\lambda\beta_{h}.
\]
 The first order condition is 
\[
\frac{\partial L\left(\beta_{h}\right)}{\partial\beta}=-\sum_{t=1}^{T}z_{t}\left(y_{t+h}-s_{t}\beta_{h}\right)+\lambda=0
\]
 such that the Lagrange multiplier, using $\beta=0,$ is 
\[
\hat{\lambda}=\sum_{t=1}^{T}z_{t}y_{t+h}.
\]
The LM test now is based on the asymptotic $\chi_{1}^{2}$ statistic
\[
T_{h}^{2}=\frac{\left(T-h\right){}^{-1}\hat{\lambda}^{2}}{Var\left(T^{-1/2}\hat{\lambda}\right)}\rightarrow_{d}\chi_{1}^{2}\text{ under \ensuremath{H_{0}}. }
\]
Let $\hat{\omega}$ be an estimator of $Var\left(n^{-1/2}\hat{\lambda}\right)$.
Then the test rejects $\beta_{h}=0$ if 
\[
\hat{T}_{h}^{2}=\frac{\left(T-h\right){}^{-1}\hat{\lambda}^{2}}{\hat{\omega}}>c_{\chi_{1}^{2},\alpha}
\]
 where $c_{\chi_{1}^{2},\alpha}$ is the $1-\alpha$ quantile of the
$\chi_{1}^{2}$ distribution. Note that 
\begin{align*}
\hat{T}_{h}^{2} & =\frac{\left(T-h\right){}^{-1}\left(\sum_{t=1}^{T}z_{t}y_{t+h}\right)^{2}}{\hat{\omega}}\\
 & =\frac{\left(T-h\right)\left(\hat{\beta}_{h}\right)^{2}}{\hat{\omega}/\left(\frac{1}{T-h}\sum_{1}^{T}z_{t}\>s_{t}\right)^{2}}
\end{align*}
which shows that the test based on $\hat{T}_{h}^{2}$ is numerically
identical to the test implied by inverting the confidence interval
proposed below. The estimator $\hat{\omega}$ can be understood as
the HAC estimator of 
\[
\left(T-h\right)^{-1}\sum_{t=1}^{T}\text{\ensuremath{z_{t}}}\tilde{u}_{t+h}=\left(T-h\right)^{-1}\sum_{t=1}^{T}z_{t}\left(y_{t+h}-s_{t}\beta_{h,0}\right)=\left(T-h\right)^{-1}\sum_{t=1}^{T}z_{t}y_{t+h}
\]
where $\left(T-h\right)^{-1}\sum_{t=1}^{T}z_{t}y_{t+h}$ is the coefficient
in an OLS regression of $z_{t}y_{t+h}$ on a constant. In other words
the variance of $\hat{\beta}_{h}$ is computed by imposing the null
when obtaining the model residual $\tilde{u}_{t+h}=y_{t+h}.$ The
long run variance (spectrum at zero frequency) of $\text{\ensuremath{z_{t}}}\tilde{u}_{t+h}=z_{t}y_{t+h}$
is $\sum_{j=-\infty}^{\infty}E(z_{t}\>y_{t+h}\>z_{t-j}\>y_{t+h-j})$
whether or not the null holds true in the DGP. The test statistic
is evaluated at the null and is based on an estimator for $\sigma^{2}=\frac{\tilde{\omega}}{\gamma_{zs}^{2}}$
where $\tilde{\omega}=\sum_{j=-\infty}^{\infty}E(z_{t}\>y_{t+h}\>z_{t-j}\>y_{t+h-j}).$ }{\small\par}

\subsection{Bonferroni Inequality}

{\small The Bonferroni inequality in the context of our impulse response
coefficients can be stated as
\begin{align*}
 & P\left(\bigcup_{h=0}^{H-1}\left\{ \hat{\beta}_{h}\notin\left[\zeta_{\alpha/2H}\,\frac{\sigma}{\sqrt{T-h}},\zeta_{(1-\alpha/2H)}\,\frac{\sigma}{\sqrt{T-h}}\right]\right\} \right)\\
 & \leq\sum_{h=0}^{H-1}P\left(\left\{ \hat{\beta}_{h}\notin\left[\zeta_{\alpha/2H}\,\frac{\sigma}{\sqrt{T-h}},\zeta_{(1-\alpha/2H)}\,\frac{\sigma}{\sqrt{T-h}}\right]\right\} \right)\\
 & \approx\sum_{h=0}^{H-1}\frac{\alpha}{H}=\alpha.
\end{align*}
where the approximation in the last line refers to the fact that the
individual confidence intervals have coverage $1-\alpha/H$ in large
samples. Based on calculations in the next section one obtains the
inequality (see also Dunn Equation 5), that the probability of all
impulse response coefficients to be inside the confidence sets is
\[
P\left(\bigcap_{h=0}^{H-1}\left\{ \zeta_{\alpha/2H}\,\frac{\sigma}{\sqrt{T-h}}<\hat{\beta}_{h}<\zeta_{(1-\alpha/2H)}\,\frac{\sigma}{\sqrt{T-h}}\right\} \right)\geq1-\alpha
\]
and where $\left\{ a<\hat{\beta}_{h}<b\right\} $ denotes the set
of all samples where $\hat{\beta}_{h}\in\left[a,b\right]$. }{\small\par}

\subsection{Joint Test of Non-Zero Impulse Responses}

{\small We define a test statistic that rejects $H_{0}:\beta=0$ if
at least one of the elements of $\beta=\left[\beta_{0},...,\beta_{H-1}\right]$
is outside the confidence interval. Formally, we say that we reject
$H_{0}$ if 
\[
T_{n}=\sum_{h=0}^{H-1}1\left\{ \hat{\beta}_{h}\notin\left[\zeta_{\alpha/2H}\hat{s}_{{\beta}_{h}}^{b},\,\,\zeta_{1-\alpha/2H}\hat{s}_{\beta_{h}}^{b}\right]\right\} >0
\]
 where $1\left\{ .\right\} $ is equal to $1$ if the argument is
true, and zero otherwise. In words, the expression counts the number
of $\hat{\beta}_{h}$ that are not inside the confidence interval.
This means that we reject the null if at least one of the estimates
is outside the confidence band around zero. The probability of rejecting
the null is 
\begin{align*}
\Pr\left(T_{n}>0\right) & =\Pr\left(\bigcup_{h=0}^{H-1}\left\{ \hat{\beta}_{h}\notin\left[\zeta_{\alpha/2H}\hat{s}_{{\beta}_{h}}^{b},\,\,\zeta_{1-\alpha/2H}\hat{s}_{\beta_{h}}^{b}\right]\right\} \right)\\
 & \leq\sum_{h=0}^{H-1}\Pr\left(\hat{\beta}_{h}\notin\left[\zeta_{\alpha/2H}\hat{s}_{{\beta}_{h}}^{b},\,\,\zeta_{1-\alpha/2H}\hat{s}_{\beta_{h}}^{b}\right]\right)\approx\sum_{h=0}^{H-1}\frac{\alpha}{H}=\alpha
\end{align*}
and where the inequality is Bonferroni's inequality. }{\small\par}

Now we construct a confidence region in $\mathbb{R}^{H}$ that contains
the estimated impulse response $\hat{\beta}$ in repeated samples
with at least probability $1-\alpha$ if the null $H_{0}$ is true.
Define the following sets
\[
A_{h}=\left\{ b=\left(b_{0},...,b_{H-1}\right)\in\mathbb{R}^{H}|b_{h}\in\left[\zeta_{\alpha/2H}\hat{s}_{{\beta}_{h}}^{b},\,\,\zeta_{1-\alpha/2H}\hat{s}_{\beta_{h}}^{b}\right]\right\} 
\]
where $A_{h}$ is a strip in $\mathbb{R}^{H}$ that goes through the
interval $\left[\zeta_{\alpha/2H}\hat{s}_{{\beta}_{h}}^{b},\,\,\zeta_{1-\alpha/2H}\hat{s}_{\beta_{h}}^{b}\right]$
on axis $h-1$. For example, for $H=2$, $A_{1}$ is the set of all
values $b=\left(b_{0},b_{1}\right)$ such that $b_{1}\in\left[\zeta_{\alpha/4}\hat{s}_{{\beta}_{h}}^{b},\,\,\zeta_{1-\alpha/4}\hat{s}_{\beta_{1}}^{b}\right]$and
$b_{0}$$\in\mathbb{R}$ is unconstrained. Then, 
\[
\bigcap_{h=0}^{H-1}A_{h}=\left\{ b\in\mathbb{R}^{H}|b_{h}\in\left[\zeta_{\alpha/2H}\hat{s}_{{\beta}_{h}}^{b},\,\,\zeta_{1-\alpha/2H}\hat{s}_{\beta_{h}}^{b}\right],h=0,...,H-1\right\} 
\]
is a rectangle in the $H$-dimensional space such that each of the
coordinates of $b$ are in the one dimensional confidence interval.
By De Morgan's law $\left(\bigcap_{h=0}^{H-1}A_{h}\right)^{c}=\bigcup_{h=0}^{H-1}A_{h}^{c}$,
where $\left(.\right)^{c}$ denotes the complement. We then have 
\begin{align}
\Pr\left(\hat{\beta}\in\left(\bigcap_{h=0}^{H-1}A_{h}\right)\right) & =1-\Pr\left(\hat{\beta}\in\left(\bigcap_{h=0}^{H-1}A_{h}\right)^{c}\right)\nonumber \\
 & =1-\Pr\left(\bigcup_{h=0}^{H-1}\left\{ \hat{\beta}_{h}\notin\left[\zeta_{\alpha/2H}\hat{s}_{{\beta}_{h}}^{b},\,\,\zeta_{1-\alpha/2H}\hat{s}_{\beta_{h}}^{b}\right]\right\} \right)\label{eq:Total-Bonf-ineq}\\
 & \geq1-\sum_{h=0}^{H-1}\Pr\left(\hat{\beta}_{h}\notin\left[\zeta_{\alpha/2H}\hat{s}_{{\beta}_{h}}^{b},\,\,\zeta_{1-\alpha/2H}\hat{s}_{\beta_{h}}^{b}\right]\right)=1-\alpha\nonumber 
\end{align}
where the first equality uses that $\left(\bigcap_{h=0}^{H-1}A_{h}\right)$
and $\left(\bigcap_{h=0}^{H-1}A_{h}\right)^{c}$ are disjoint, and
$\left(\bigcap_{h=0}^{H-1}A_{h}\right)\cup\left(\bigcap_{h=0}^{H-1}A_{h}\right)^{c}=\mathbb{R}^{H}.$
The second equality uses 
\[
\left\{ \hat{\beta}\in\left(\bigcap_{h=0}^{H-1}A_{h}\right)\right\} =\left(\bigcap_{h=0}^{H-1}\left\{ \hat{\beta}\in A_{h}\right\} \right)
\]
since $\hat{\beta}$ is in the hypercube $\left(\bigcap_{h=0}^{H-1}A_{h}\right)$
if all coordinates $\hat{\beta}_{h}$ are in the segments defining
the hypercube. Here the $\left\{ .\right\} $ brackets mean all outcomes
for which $\hat{\beta}$ satisfies the condition. Since $\hat{\beta}\in A_{h}$
iff $\hat{\beta}_{h}\in\left[\zeta_{\alpha/2H}\hat{s}_{{\beta}_{h}}^{b},\,\,\zeta_{1-\alpha/2H}\hat{s}_{\beta_{h}}^{b}\right]$
it follows that $\left\{ \hat{\beta}\in A_{h}\right\} ^{c}=\left\{ \hat{\beta}_{h}\notin\left[\zeta_{\alpha/2H}\hat{s}_{{\beta}_{h}}^{b},\,\,\zeta_{1-\alpha/2H}\hat{s}_{\beta_{h}}^{b}\right]\right\} $
Now apply de Morgan's law to the RHS. Finally, the inequality in the
display above is the Bonferroni inequality. 

Since we do not reject $H_{0}$ if $T_{n}=0$ and $T_{n}=0$ iff $\hat{\beta}_{h}\in\left[\zeta_{\alpha/2H}\hat{s}_{{\beta}_{h}}^{b},\,\,\zeta_{1-\alpha/2H}\hat{s}_{\beta_{h}}^{b}\right]$
for all $h=0,...,H-1$ we obtain 
\begin{align*}
\Pr\left(T=0\right) & =\Pr\left(\bigcap_{h=0}^{H-1}\left\{ \hat{\beta}_{h}\in\left[\zeta_{\alpha/2H}\hat{s}_{{\beta}_{h}}^{b},\,\,\zeta_{1-\alpha/2H}\hat{s}_{\beta_{h}}^{b}\right]\right\} \right)\\
 & =1-\Pr\left(\bigcup_{h=0}^{H-1}\left\{ \hat{\beta}_{h}\notin\left[\zeta_{\alpha/2H}\hat{s}_{{\beta}_{h}}^{b},\,\,\zeta_{1-\alpha/2H}\hat{s}_{\beta_{h}}^{b}\right]\right\} \right)\\
 & \geq1-\alpha
\end{align*}
where the second equality follows from the laws of total probability
and De Morgan's law and the inequality follows from the bound in \eqref{eq:Total-Bonf-ineq}. 

\end{document}